# On Stress-Strain Responses and Young's Moduli of Single Alkane Molecules, A Molecular Mechanics Study Using the Modified Embedded-Atom Method

*Sasan Nouranian,*[*a] *Steven R. Gwaltney,*[b] *Michael I. Baskes,*[c] *Mark A. Tschopp,*[d] *Mark F. Horstemeyer*[ae]

[a] Center for Advanced Vehicular Systems (CAVS), Mississippi State University, Mississippi State, MS 39762, USA

[b] Department of Chemistry, Mississippi State University, Mississippi State, MS 39762, USA

[c] Department of Aerospace Engineering, Mississippi State University, Mississippi State, MS 39762, USA

[d] U.S. Army Research Laboratory, Aberdeen Proving Ground, MD 21005, USA

[e] Department of Mechanical Engineering, Mississippi State University, Mississippi State, MS 39762, USA

**ABSTRACT**

The stress-strain response of a single polymer chain has hitherto remained elusive due to the fact that the cross-sectional area and volume of a single molecule are not well-defined. However, a knowledge of this mechanical response is both essential and a pre-requisite to understanding the nanomechanics of fracture toward developing a multi-scale model for damage prediction in polymers. In this work, molecular mechanics simulations were performed using a modified embedded-atom method (MEAM) potential to generate the stress-strain responses of a series of *n*-alkane molecules from ethane ($C_2H_6$) to undecane (*n*-$C_{11}H_{24}$) in tensile deformation up to the point of bond rupture. The results are further generalized to a single polyethylene (PE) chain. Force, true Cauchy stress, true virial stress, and Young's moduli were calculated as a function of true strain for all of the molecules. In calculating the stress of a single molecule, three methods (designated in this work as M1, M2, and M3) are suggested based on three different metrics to quantify both the instantaneous molecular cross-sectional area and volume during deformation. The predictions of these methods are compared to theoretical, first-principles, and experimental data. M1 gives true Cauchy and true virial stress results that are essentially equivalent, suggesting that it is a better method for calculating stress in alkane molecules and, hence, PE single chain. The MEAM predictions of the average elastic modulus for a single PE chain using M1, M2, and M3 are 401 GPa, 172 GPa, and 147 GPa, respectively. Though these results are disparate, M1 gives a modulus value that is strikingly close to the *ab initio*-calculated value of 405 GPa for the -$CH_2CH_2$- repeat unit of PE at 0 K. To further test the MEAM potential, the Young's modulus ($C_{33}$ elastic constant) of an orthorhombic PE crystal cell was calculated, but its value (184 GPa) underestimates the DFT-calculated value of 316 GPa by 42%. Overall, the MEAM potential gives reasonable predictions of the nanomechanics of bond rupture and very good predictions of the Young's modulus of a single PE chain. However, it underestimates the modulus of a PE crystal.

---

[*] Present address: Department of Chemical Engineering, The University of Mississippi, University, MS 38677, USA. Email: sasan@olemiss.edu (S. N.)

## 1. Introduction

Molecular bond rupture under tensile deformation is of great fundamental interest for damage and failure characterization and modeling of polymers from both mechanochemistry and molecular nanomechanics points of view [1-3]. An understanding and quantification of tensile bond rupture energetics and mechanisms, molecular fragment formation, and stress-strain response of a single molecule provides a basis for developing predictive multi-scale damage models for polymers that ultimately incorporate void formation, growth, and coalescence [4]. While the significance of determining the single-chain stress-strain response is evident from the above discussion, it has remained a point of controversy since the effective cross-sectional area and/or volume of a single molecule for stress calculation is not well-defined. There are numerous theoretical [5, 6], first-principles (FP) [7-12], and experimental data [13-15] available for the Young's modulus of a single PE chain, in which various effective chain cross-sectional area are used for the modulus calculation. Furthermore, the reported data are either for an isolated all-trans PE chain or a chain in the crystalline environment. However, the effect of crystalline environment on the modulus of a single PE chain is minimal [15]. The disparity in reported Young's modulus values for a single PE chain compels one to exercise caution when comparing the different methods.

One key technique far less used in studying and quantifying the nanomechanics of single polymer chains is molecular mechanics simulation using reactive interatomic potentials. This technique has emerged as an effective tool for studying the mechanochemistry of molecular fracture under tensile deformation [16]. In our previous study [16], we presented the energetics and fragment formation investigation of a series of *n*-alkane molecules using molecular statics calculations with our recently parameterized reactive modified embedded-atom method (MEAM) potential for saturated hydrocarbons [17]. We further generalized the results to a single PE chain. In this study, we investigate the nanomechanics of the above saturated hydrocarbon series and quantify their stress-strain response. We further determine the Young's modulus of a single PE chain. Since stress definitions have yet not been analyzed at these small length scales in polymers, we employ different definitions in this work and analyze the predictions relative to higher length scales. A similar notion of analyzing stress definitions was successfully accomplished by Horstemeyer et al. [18, 19]. We also calculate the bulk Young's modulus of an orthorhombic PE crystal cell using the MEAM potential and compare the results to the experimental data. One objective of the current work is to introduce proper methods for determining the instantaneous cross-sectional area of a single hydrocarbon molecule during molecular deformation toward investigating molecular nanomechanics of fracture. By correlating the calculated cross-sectional areas and the predicted moduli of a single PE chain to the *ab initio* calculated and experimentally measured values, we provide a comparison basis for future molecular mechanics investigations of fracture in polymers.

## 2. Computational method

All computations were performed in the open-source large-scale atomic/molecular massively parallel simulator (LAMMPS) software package [20] utilizing its built-in MEAM potential file. The carbon, hydrogen, and CH parameters for saturated hydrocarbons were published previously [17] and can be accessed through the NIST Interatomic Potentials Repository Project website at http://www.ctcms.nist.gov/potentials/C-H.html. Alkane molecules from ethane ($C_2H_6$) to undecane (*n*-C$_{11}$H$_{24}$) were created in all-*trans* configurations and then subjected to a stepwise elongation process that is described in detail in our previous work [16]. We reiterate here that the coordinates of the terminal carbon atom for the molecules with even numbers of carbon atoms in their backbone were zeroed in the first step of the deformation. Force versus true strain and true stress versus true strain behaviors were quantified for the different molecules. Calculating the stress for single molecules is not trivial but very important in the context of multiscale materials modeling. In this work, different methods were used to calculate the true Cauchy and true virial stress. The main difference between the three methods is how to define the length, cross-sectional area, and volume of the single alkane molecules. Calculating the stress-strain response of single alkane molecules is a necessary step for calculating the stiffness (Young's modulus) of longer chain molecules, *i.e.*, PE, and incorporating bond rupture into higher scale formulations for fracture of polymers. In particular, macroscale continuum models use the Cauchy stress, since it the stress metric used in the conservation laws of mass, momentum (linear and angular), and energy; hence, it is important to reconcile potential differences between the Cauchy and virial stress formulations at the molecular level, as well as any differences due to length/area/volume calculations within a molecule. The aforementioned three different methods are described in the next subsections.

**2.1. Method 1 for true stress calculation (quadrilateral area)**
In the first method (M1), the two upper and lower hydrogen atoms situated on the second and the next-to-terminal carbon atoms are taken as the four corners of a quadrilateral seen from the front or back of the molecule (Fig. 1-a). The force at each increment step is divided by the instantaneous area of this quadrilateral to give the true Cauchy stress. To calculate the true virial stress, the *xx*-component of the stress×volume tensor at each increment step is divided by the instantaneous volume calculated by multiplying the above-mentioned area by the instantaneous length of the molecule determined from the coordinates of the first and terminal carbon atoms. The true strain calculation in this method utilizes the above-mentioned procedure to determine the instantaneous length of the molecule.

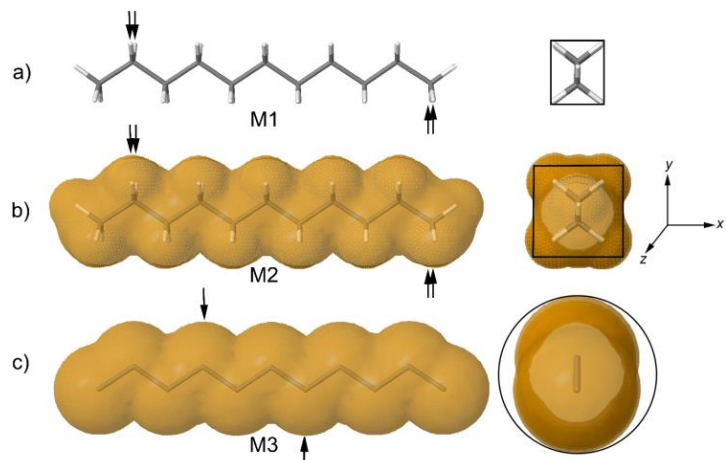

**Fig. 1** Front and right-side views of an alkane molecule (C$_{11}$H$_{24}$), where the cross-sectional area calculated using a) M1, b) M2, and c) M3 are shown schematically. The arrows point to the possible coordinates of the hydrogen atoms comprising the corners of the quadrilaterals in a) and b), and the carbon atoms comprising the diameter of the cricle in c). In c), only carbon atoms are shown.

**2.2. Method 2 for true stress calculation (van der Waals corrected quadrilateral area)**
In the second method (M2), the same procedure is followed as described for M1, except the corners of the quadrilateral are determined from the coordinates of the hydrogen atoms corrected by the van der Waals (vdW) radius of atomic hydrogen, which is taken to be 1.2 Å [21]. The correction is made along the diagonals of the quadrilateral (Fig. 1-b). Similarly, the volume is determined by multiplying the instantaneous vdW-corrected area by the instantaneous length of the molecule, corrected for the vdW radius of atomic carbon, which is taken to be 1.7 Å [22]. The correction is made on both sides along the line connecting the initial and terminal carbon atoms. As an example, M2 yields a cross-sectional area for the minimum molecular configuration of *n*-decane that is roughly three times larger and a volume that is roughly four times larger than those calculated using M1. Both the Cauchy and virial stresses can be calculated using M2 as well. The true strain calculation in this method utilizes the same vdW-corrected instantaneous length of the molecule.

**2.3. Method 3 for true stress calculation (circular area)**
In the third method (M3), the circular instantaneous cross-sectional area of the molecule is calculated by defining a vdW-corrected diameter comprised of the *y*-distance between two outwardly located carbon atoms (Fig. 1-c) on either side of the molecular backbone. The instantaneous molecular volume is calculated, similar to M2, by multiplying the circular area at each increment step by the vdW-corrected length of the molecule. M3 yields a cross-sectional area and volume for the minimum configuration of the *n*-decane molecule that are both roughly 30% larger than the values calculated using M2 for the Cauchy and virial stresses. The true strain calculation in this method is the same as that for M2.

A comparison between the calculated cross-sectional areas of *n*-decane and *n*-undecane molecules using M1, M2, and M3 and the experimentally determined value of the area for a PE chain is given in Table 1. We note here that zeroing of the terminal carbon atom coordinates in even carbon-numbered alkanes (for example *n*-decane in Table 1) before deformation causes a slight titling of the molecule and, hence, the calculated areas using the three methods (M1, M2, and M3) become larger than those calculated for the odd carbon-numbered alkanes (for example *n*-undecane in Table 1).

**Table 1** Cross-sectional areas (Å$^2$) of *n*-decane and *n*-undecane molecules calculated using M1, M2, and M3 versus the experimental value for a polyethylene (PE) chain.

| Molecule | M1 | M2 | M3 | Expt. for PE [23] |
|---|---|---|---|---|
| $C_{10}H_{22}$ | 5.0 | 14.8 | 19.1 | 18.24 |
| $C_{11}H_{24}$ | 3.8 | 13.2 | 14.4 | |

## 3. Results and discussion
The true stress versus true strain for the *n*-alkane molecules from $C_3H_8$ (propane) to $C_{11}H_{24}$ (*n*-undecane) are given in Figs. 2 and 3 for the Cauchy and virial stress, respectively. The results are generalized to a single PE chain since higher alkanes, such as undecane, can represent the PE chain adequately [24]. Both Cauchy and virial stress data were calculated based on the three area methods described in detail in the previous section. In general, the molecule's constitutive behavior overlaps separately for higher alkanes with odd and even numbers of carbon atoms in the molecular backbone. As mentioned before, this observation can be attributed to the slight difference in the cross-sectional area calculated for the molecules with odd and even numbers of carbon atoms in their backbone. An average between the two can be calculated, and the resulting true stress versus true strain can be generalized for a longer single PE chain. M1 (Method 1) yields a much higher ultimate tensile strength for $C_{10}H_{22}$ than that calculated using M2 and M3 (Fig. 4). Furthermore, the elongation at fracture is shorter when calculated using M2 and M3. This observation has to do with the vdW correction applied when calculating the true strain (see the previous section). Another important observation is that the true Cauchy and true virial stress versus true strain coincide when M1 is used (Fig. 4). However, there is an 18% difference between the true Cauchy and true virial stress calculated using M2 and M3 at longer true strains. Despite some claims that the atomistic virial stress calculation is not equivalent to the continuum Cauchy stress, Subramaniyan and Sun [25] show that the virial stress is indeed equivalent to the Cauchy stress. Therefore, the near-equivalence of the true Cauchy and true virial stress curves with M1 suggests that M1 is a better method for calculating the stress in alkane molecules and, hence, PE single chains. As shown in Fig. 4, the Cauchy and virial stress-strain curves for M2 and M3 are essentially equivalent. This observation can be attributed to the fact that carbon has a larger van der Waals radius than hydrogen (1.7 Å versus 1.2 Å, respectively) and the diameter for the cross-sectional area calculated in M3 (for example, 4.94 Å for *n*-decane in its minimum configuration) is very close to the size of the quadrilaterial side calculated in M2 (roughly 5.07 Å for *n*-decane in its minimum configuration).

The MEAM potential predicts an estimated ultimate tensile strength of 83-85 GPa for a single PE chain when an average is calculated (based on either Cauchy or virial stress and M1) for the *n*-alkanes with odd and even number of carbon atoms in their molecular backbone (Figs. 2a and 3a). It is again assumed that these values extrapolate to longer hydrocarbon chains. Brower *et al.* [26] predict an ultimate tensile strength of 66 GPa for the –$CH_2CH_2$- repeat unit of PE at 0 K. If this value is taken as an estimate of the ultimate tensile strength of a single PE chain, the MEAM prediction using M1 is within 26% of the FP prediction. Hageman *et al.* [27] used *ab initio* calculations of the bond scission rate in a PE chain and reported an ultimate tensile strength of 18 GPa at an engineering strain of 8% (true strain of 7.7%). They used a room-temperature cross-sectional area of 18.24 Å$^2$ for a PE chain [23] to calculate the tensile strength. In this case, the MEAM predictions of the ultimate tensile strength for the *n*-decane molecule calculated using methods M2 and M3 in the true virial stress versus true strain (Figs. 3b-3c) are 19.4 GPa and 17.8 GPa, respectively. In these two methods, the average initial undeformed cross-sectional area of the *n*-decane molecule is ~15-19 Å$^2$ (Table 1), which is close to the cross-sectional area used by Hageman *et al.* (18.24 Å$^2$) [23]. The MEAM-predicted ultimate strength values could be used as estimated values for a single PE chain by the same arguments as those discussed above. However, the elongation at fracture predicted by the MEAM potential (28% for M1) is significantly larger than that reported by Hageman *et al.* (7.7%).

Another important mechanics parameter is the Young's modulus, which was calculated for the alkane molecules from $C_5H_{12}$ to $C_{11}H_{24}$ based on the initial slope of the curves presented in Figs. 2 and 3. The smaller alkane molecules were left out from the analysis, since overlapping of the true stress versus true strain curves (Figs. 2 and 3) occur in higher alkanes starting from $C_5H_{12}$. The results for the Young's moduli obtained using all three methods (M1, M2, and M3) are shown in Table 2. Again, the alkane molecules with odd and even numbers of carbon atoms in their molecular backbone tend to give similar moduli, respectively. The Young's modulus is found to be fairly constant for the different *n*-alkane molecules. This observation is attributed to the fact that the modulus is dominated by the C-C-C angle bending during the initial stages of the molecular deformation [16]. However, the stress gets distributed in the chain at each strain increment causing more stretched C-C bonds towards the ends of the molecule. The contribution of C-C-C angle bending gets smaller and the C-C bond stretching gets larger as the deformation moves towards the rupture event.

Next, an average Young's modulus was calculated for each method, representing an estimated Young's modulus for a single PE chain. These averages together with the standard deviations are given in Table 2. In general, M1 yields an average modulus that is about twice the values calculated using M2 and M3. In the literature, a range of experimental and FP values are reported for the Young's modulus of a single PE chain. From FP calculations, Crist *et al*. [12] report a modulus of 405 GPa for the –$CH_2CH_2$- repeat unit of PE at 0 K, which is strikingly close to the MEAM-predicted average value of 401 GPa using M1 (Table 2). Suhai [7] reports values of 297-493.5 GPa for the longitudinal elastic modulus of PE calculated using various quantum mechanical methods. Meier [15] employed a semiempirical method to calculate a Young's modulus of 353 GPa for a single PE chain. On the experimental side, Du *et al*. [14] used nanoindentation and scanning force microscopy to measure the PE modulus along the chain axis utilizing two different methods and reported values of 168 GPa and 204 GPa for thinner specimens and 267 GPa and 278 GPa for thicker specimens. In short, the range of Young's moduli calculated or measured for a single PE chain is anywhere from 168 GPa to 493.5 GPa. Interestingly, M1 results are comparable to the FP-computed values, and M2 and M3 results are comparable to the experimental values. The latter observation can be attributed to the fact that the molecular cross-sectional areas calculated in M2 (~15 Å$^2$ for relaxed *n*-decane) and M3 (~19 Å$^2$) are close to the experimental value of 18.24 Å$^2$ [23].

**Table 2** Young's moduli (GPa) for *n*-alkane molecules calculated using M1, M2, and M3. The averages and standard deviations represent the estimated values for a single polyethylene chain.

| Molecule | M1 | M2 | M3 |
|---|---|---|---|
| $C_5H_{12}$ | 378 | 179 | 165 |
| $C_6H_{14}$ | 400 | 199 | 160 |
| $C_7H_{16}$ | 400 | 166 | 153 |
| $C_8H_{18}$ | 385 | 174 | 139 |
| $C_9H_{20}$ | 422 | 163 | 145 |
| $C_{10}H_{22}$ | 378 | 161 | 128 |
| $C_{11}H_{24}$ | 444 | 161 | 141 |
| Average | 401 | 172 | 147 |
| St. Dev. | 25 | 14 | 13 |

In addition to the calculation of Young's modulus for a single PE chain, Young's modulus of an orthorhombic PE crystal cell was also calculated in this work to further validate the MEAM potential predictions. An orthorhombic PE cell with unit cell lattice constants of $a = 7.39$ Å, $b = 4.93$ Å, and $c = 2.54$ Å was constructed, and the $C_{33}$ elastic constant was calculated by an axial deformation of the cell. The MEAM-calculated value of the $C_{33}$ elastic constant (Young's modulus) at 0 K is 184 GPa, which is an underestimation by 42% of the DFT-calculated value of 316 GPa by Lacks and Rutledge [28]. A value of 334 GPa was reported by Hageman *et al*. [9], while Barrera *et al*. [10] reported a value of 360.2 GPa for the Young's modulus of a PE crystal cell at 0 K. Nevertheless, the MEAM potential minimized the cell to the correct lattice parameters.

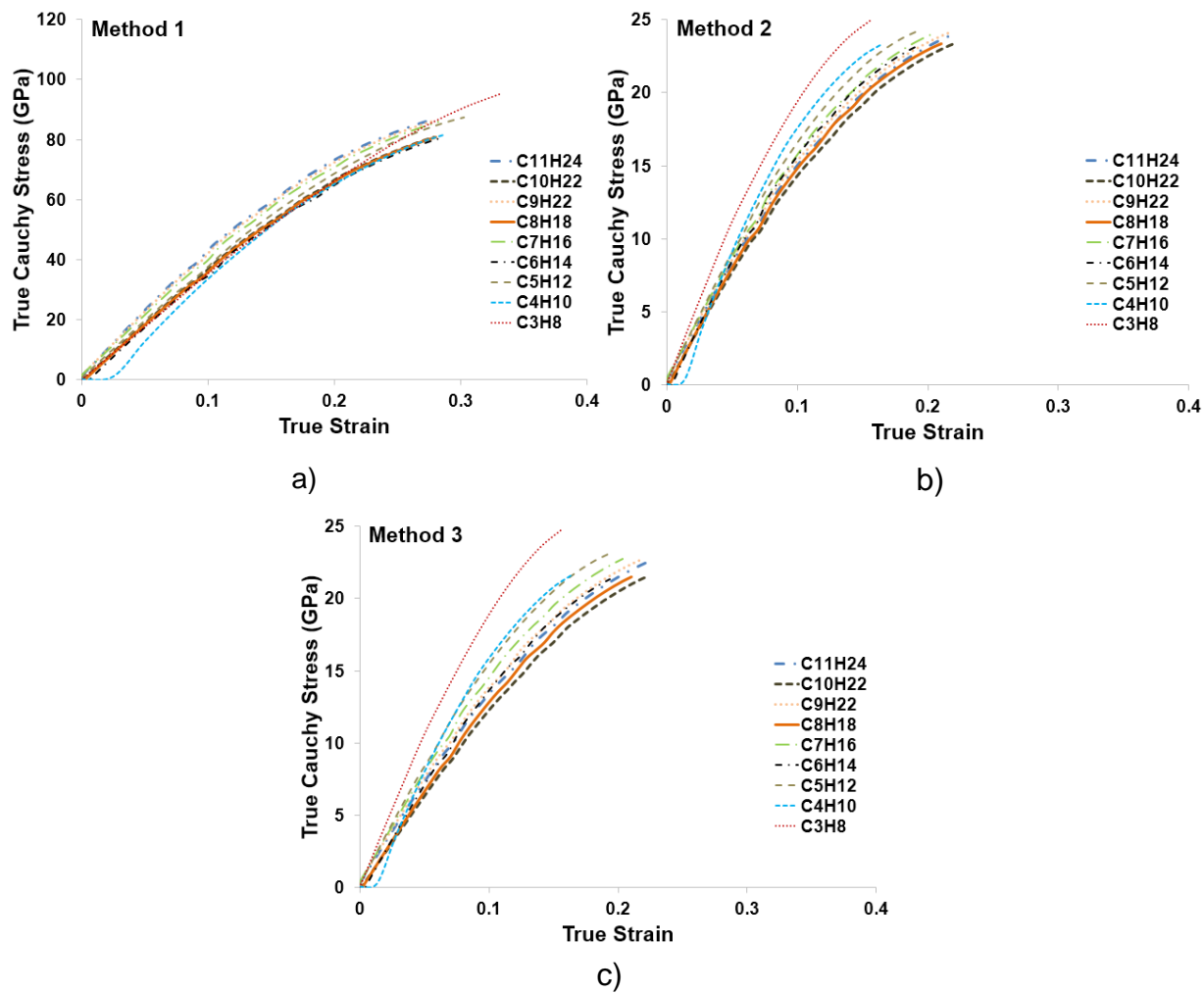

**Fig. 2** True Cauchy stress versus true strain for a series of alkane molecules from $C_3H_8$ to $C_{11}H_{24}$ calculated using (a) Method 1 (M1), (b) Method 2 (M2), and (c) Method 3 (M3). These methods are described in the Computational Method section.

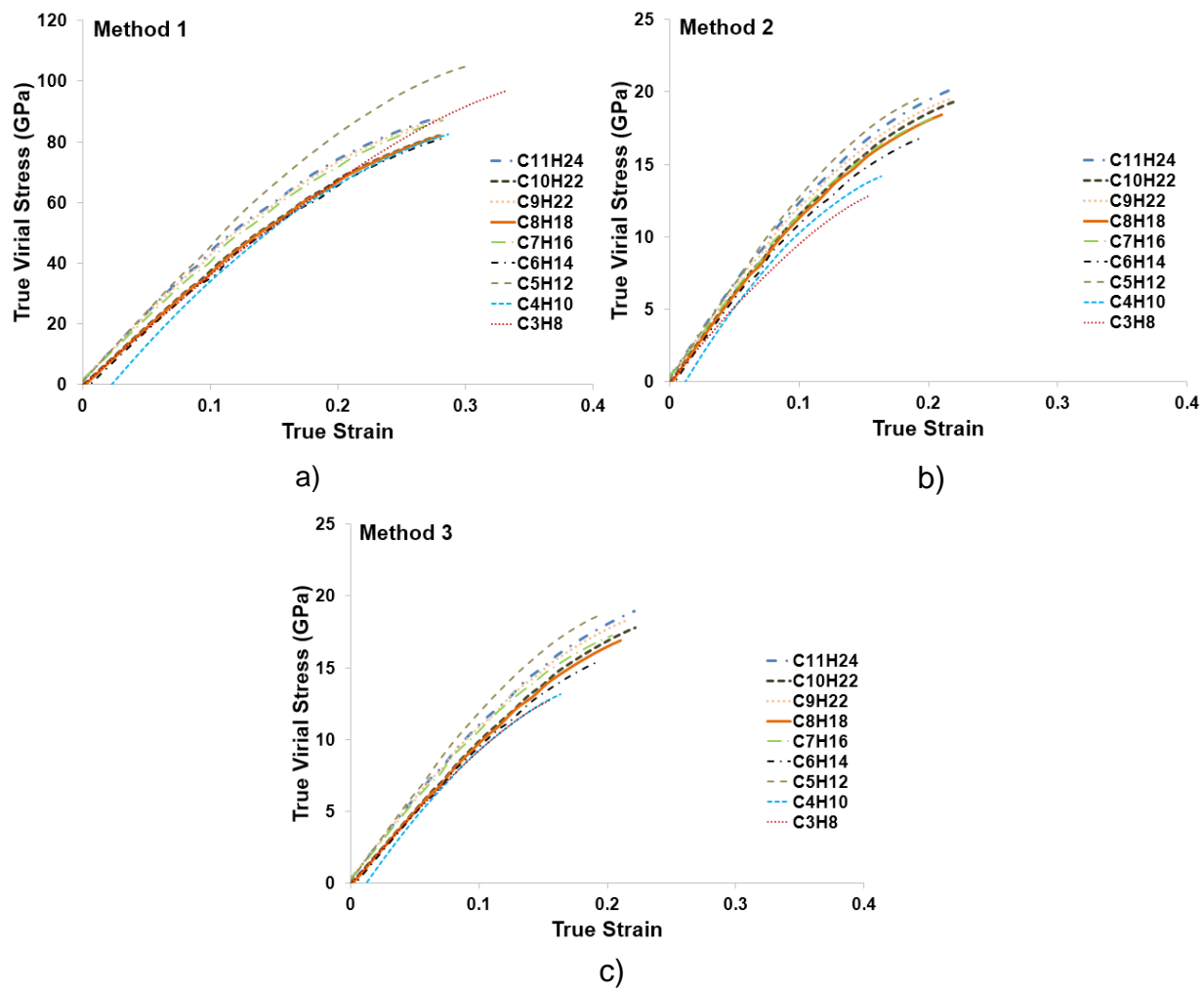

**Fig. 3** True virial stress versus true strain for a series of alkane molecules from $C_3H_8$ to $C_{11}H_{24}$ calculated using (a) Method 1 (M1), (b) Method 2 (M2), and (c) Method 3 (M3). These methods are described in the Computational Method section.

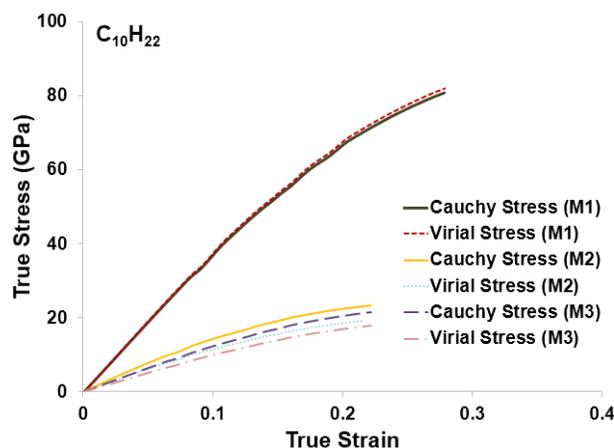

**Fig. 4** Comparisons between the true Cauchy and true virial stress calculated using three methods as a function of true strain for the *n*-decane ($C_{10}H_{22}$) molecule. M1, M2, and M3 denote Method 1, Method 2, and Method 3, respectively. Note that M1 gives the greatest "hardening" rate.

## 4. Summary and conclusions

With respect to a single PE chain, the calculated ultimate tensile strength using one of the methods for determining the stress response (Method 1) is within 26% of the FP prediction. Using the same method, an average Young's modulus of 401 GPa is calculated for a single PE chain, which is remarkably within 1% of the FP results for the $-CH_2CH_2-$ repeat unit of PE at 0 K. However, the calculated Young's modulus for an orthorhombic PE crystal cell using the MEAM potential (184 GPa) underestimates the DFT-calculated value (316 GPa) by 42%.

In this work, different methods of quantifying the cross-sectional area to define the true Cauchy stress were introduced, and it was shown that the correlated virial stress gives very close results in these inelastic deformations. The stress and moduli revealed in these molecular statics simulations give fairly close results to the experimental data and FP results.

## Acknowledgments

This work was sponsored by the U.S. Army Engineer Research and Development Center (ERDC) under contract no. W15QKN-13-9-0001. Special thanks go to Robert Moser for his support.

## Distribution Statement

Approved for public release; distribution unlimited.

## References

[1]  S. N. Zhurkov and V. E. Korsukov, "Atomic mechanism of fracture of solid polymers," *Journal of Polymer Science: Polymer Physics Edition,* vol. 12, pp. 385-398, 1974.

[2]  M. K. Beyer and H. Clausen-Schaumann, "Mechanochemistry: The Mechanical Activation of Covalent Bonds," *Chemical Reviews,* vol. 105, pp. 2921-2948, 2005/08/01 2005.

[3]  W. Zhang and X. Zhang, "Single molecule mechanochemistry of macromolecules," *Progress in Polymer Science,* vol. 28, pp. 1271-1295, 8// 2003.

[4] D. K. Francis, J. L. Bouvard, Y. Hammi, and M. F. Horstemeyer, "Formulation of a damage internal state variable model for amorphous glassy polymers," *International Journal of Solids and Structures,* vol. 51, pp. 2765-2776, 8/1/ 2014.

[5] L. R. G. Treloar, "Calculations of elastic moduli of polymer crystals: I. Polyethylene and nylon 66," *Polymer,* vol. 1, pp. 95-103, // 1960.

[6] I. M. Ward and J. Sweeney, *Mechanical Properties of Solid Polymers*: Wiley, 2012.

[7] S. Suhai, "Quantum mechanical calculation of the longitudinal elastic modulus and of deviations from Hooke's law in polyethylene," *The Journal of chemical physics,* vol. 84, pp. 5071-5076, 1986.

[8] K. Palmö and S. Krimm, "Chain elastic modulus of polyethylene: A spectroscopically determined force field (SDFF) study," *Journal of Polymer Science Part B: Polymer Physics,* vol. 34, pp. 37-45, 1996.

[9] J. Hageman, R. J. Meier, M. Heinemann, and R. De Groot, "Young modulus of crystalline polyethylene from ab initio molecular dynamics," *Macromolecules,* vol. 30, pp. 5953-5957, 1997.

[10] G. D. Barrera, S. F. Parker, A. J. Ramirez-Cuesta, and P. C. Mitchell, "The vibrational spectrum and ultimate modulus of polyethylene," *Macromolecules,* vol. 39, pp. 2683-2690, 2006.

[11] A. Karpfen, "Abinitio studies on polymers. V. All-trans-polyethylene," *The Journal of Chemical Physics,* vol. 75, pp. 238-245, 1981.

[12] B. Crist, M. Ratner, A. Brower, and J. Sabin, "Abinitio calculations of the deformation of polyethylene," *Journal of Applied Physics,* vol. 50, pp. 6047-6051, 1979.

[13] T. N. Wassermann, J. Thelemann, P. Zielke, and M. A. Suhm, "The stiffness of a fully stretched polyethylene chain: a Raman jet spectroscopy extrapolation," 2009.

[14] B. Du, J. Liu, Q. Zhang, and T. He, "Experimental measurement of polyethylene chain modulus by scanning force microscopy," *Polymer,* vol. 42, pp. 5901-5907, 2001.

[15] R. J. Meier, "Regarding the ultimate Young's modulus of a single polyethylene chain," *Macromolecules,* vol. 26, pp. 4376-4378, 1993.

[16] S. Nouranian, S. R. Gwaltney, M. I. Baskes, M. A. Tschopp, and M. F. Horstemeyer, "Simulations of Tensile Bond Rupture in Single Alkane Molecules Using Reactive Interatomic Potentials," *Chemical Physics Letters,* vol. 635, pp. 278-284, 2015.

[17] S. Nouranian, M. A. Tschopp, S. R. Gwaltney, M. I. Baskes, and M. F. Horstemeyer, "An Interatomic Potential for Saturated Hydrocarbons Based on the Modified Embedded-Atom Method," *Physical Chemistry Chemical Physics,* vol. 16, pp. 6233-6249, 2014.

[18] M. F. Horstemeyer and M. Baskes, "Atomistic finite deformation simulations: a discussion on length scale effects in relation to mechanical stresses," *Journal of engineering materials and technology,* vol. 121, pp. 114-119, 1999.

[19] M. F. Horstemeyer, M. I. Baskes, and S. J. Plimpton, "Computational nanoscale plasticity simulations using embedded atom potentials," *Theoretical and Applied Fracture Mechanics,* vol. 37, pp. 49-98, 2001.

[20] S. Plimpton, "Fast parallel algorithms for short-range molecular dynamics," *Journal of Computational Physics,* vol. 117, pp. 1-19, 1995.

[21] G. Raj, *Advanced Inorganic Chemistry Vol-1*: Krishna Prakashan.

[22] M. M. Deza and E. Deza, *Encyclopedia of Distances*: Springer, 2009.

[23] C. W. Bunn, *Transactions of the Faraday Society,* vol. 35, p. 482, 1939.

[24] A. M. Saitta and M. L. Klein, "Polyethylene under tensile load: Strain energy storage and breaking of linear and knotted alkanes probed by first-principles molecular dynamics calculations," *The Journal of chemical physics,* vol. 111, p. 9434, 1999.

[25] A. K. Subramaniyan and C. T. Sun, "Continuum interpretation of virial stress in molecular simulations," *International Journal of Solids and Structures,* vol. 45, pp. 4340-4346, 7// 2008.

[26] A. Brower, J. Sabin, B. Crist, and M. Ratner, "Ab initio molecular orbital studies of polyethylene deformation," *International Journal of Quantum Chemistry,* vol. 18, pp. 651-654, 1980.

[27] J. Hageman, G. de Wijs, R. de Groot, and R. J. Meier, "Bond scission in a perfect polyethylene chain and the consequences for the ultimate strength," *Macromolecules,* vol. 33, pp. 9098-9108, 2000.

[28] D. J. Lacks and G. C. Rutledge, "Simulation of the temperature dependence of mechanical properties of polyethylene," *The Journal of Physical Chemistry,* vol. 98, pp. 1222-1231, 1994/01/01 1994.